\documentclass{article}
\usepackage{graphicx} 
\usepackage{fullpage}
\usepackage{amsmath}
\usepackage{amsthm}
\usepackage{amssymb}
\usepackage{amsfonts,mathrsfs}
\usepackage{amsxtra}
\usepackage[switch*]{lineno}
\usepackage[citestyle=numeric-comp,sorting=none]{biblatex}
\usepackage[breaklinks]{hyperref}
\bibliography{references}

\begin{document}

\normalsize
 
\begin{center}
\vspace*{2.5cm}
{\Large
\bf
 
Review of hadronic vacuum polarization calculations via $e^+e^-$ measurements}
 
\vspace{1.cm}
{\large Zhiqing Zhang\footnote{Presented at the 32nd International Symposium on Lepton Photon Interactions at High Energies, Madison, Wisconsin, USA, August 25-29, 2025}}\\
{\small IJCLab, IN2P3/CNRS and University Paris-Saclay}
 
\end{center}
 
\date{\today}

\vspace{1.5cm}
 
\begin{abstract}

The discrepancy on the muon anomalous magnetic moment values obtained via a direct measurement and via a data-driven theory determination that uses the experimentally measured 
 hadronic cross section, is among the long standing and most significant deviations from the Standard Model predictions. The recently presented final result of the direct measurement performed at the experiment at Fermilab, with an impressive accuracy of 127 parts-per-billion, further stresses the need for a theory estimate of comparable accuracy.
The $e^+e^-$ hadronic cross section is the experimental input to the dispersive integral for the calculation of the hadronic contribution to the $g-2$, which is largely dominated by the $e^+e^-\to \pi^+\pi^-$ channel. Precise measurements of the $e^+e^-\to \pi^+\pi^-$ cross section with a sub-percent accuracy, in the energy region of the $\rho$-resonance peak, have been performed by several experiments, but the results differ way more than the published accuracies limiting the comparison with the Fermilab direct measurement.
New results are expected to become available in the near future from KLOE, CMD-3, SND and BESIII experiments, while a new measurement from the BABAR collaboration is presented today and its impact on the present situation is discussed. This measurement makes use of the entire BABAR data set and adopts a different analysis strategy, with results (still preliminary) largely independent of the results published in the 2009 BABAR publication.    
\end{abstract}

\clearpage
\section{Introduction}

For a charged lepton $\ell$ with charge $q_\ell$ and mass $m_\ell$, its magnetic moment $\vec{\mu}_\ell$ is connected to its spin $\vec{S}_\ell$ by a $g_\ell$ factor as $\vec{\mu}_\ell=g_\ell\left(\frac{q_\ell}{2m_\ell}\right)\vec{S}_\ell$.
The $g_\ell$ factor predicted by Dirac about a century ago has a value of 2~\cite{dirac1928quantum}. 
Quantum corrections of the standard model (SM) from all three sectors modify the value of $g_\ell$ by 2 per mil.
The magnetic anomaly $a_\ell$ was introduced to quantify the deviation from 2:
\begin{equation}
a_\ell=\frac{g_\ell-2}{2}\,.
\end{equation}
The muon magnetic anomaly, $a_\mu$, is one of the most precisely predicted and measured observables in particle physics.

\begin{figure}[hbtp]
    \centering
    \includegraphics[width=0.6\linewidth]{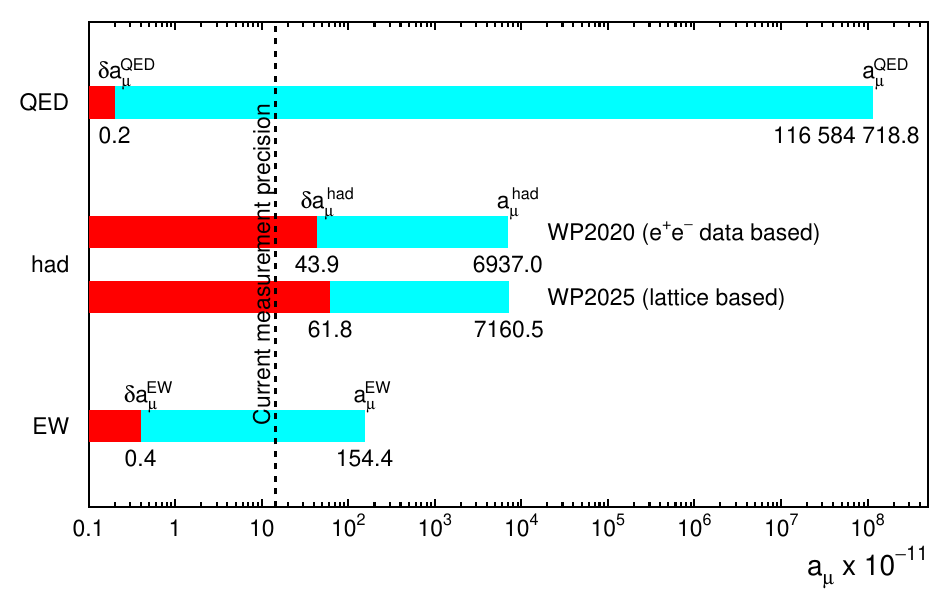}
    \caption{\small Relative contributions of the QED, hadronic (had) and EW predictions (blue bars) and their uncertainties (red bars) in comparison with the current measurement precision (dashed vertical line).}
    \label{fig:amu_sm}
\end{figure}

In the SM, $a_\mu$ receives contributions from the QED, strong and electroweak (EW) sectors:
$a_\mu^\mathrm{SM}=a_\mu^\mathrm{QED}+a_\mu^\mathrm{had}+a_\mu^\mathrm{EW}$.
The QED contribution is by far the largest one~\cite{Aoyama:2020ynm,Aliberti:2025beg}. 
Its value is predicted to such an extremely high precision that it is often shown in units of $10^{-11}$ or $10^{-10}$. 
The EW contribution is the smallest one and also well known. 
The hadronic contribution from the strong sector is between the two and has the largest uncertainty.
This is why it has been the focus of the studies since decades. 
The relative contributions of the QED, strong and EW predictions and their uncertainties are shown in Figure~\ref{fig:amu_sm}.
Two versions of the hadronic contribution are shown, one is based on $e^+e^-$ data from white paper (WP) 2020~\cite{Aoyama:2020ynm} of the muon $g-2$ theory initiative and the other is based on lattice calculations from WP 2025~\cite{Aliberti:2025beg}.
The hadronic contribution can be further decomposed into leading order (LO) hadronic vacuum polarization (HVP), higher order (HO) HVP and hadronic light-by-light (HLbL) one as:
$a_\mu^\mathrm{had}=a_\mu^\text{HVP LO}+a_\mu^\text{HVP HO}+a_\mu^\mathrm{HLbL}$.
The LO HVP contribution in turn has the largest value and uncertainty, as shown in Table~\ref{tab:amu_had}.
\begin{table}[htbp]
    \centering
    \caption{\small Predictions for three components of the hadronic contributions and their uncertainties.}
    \label{tab:amu_had}
    \vspace{2mm}
    \begin{tabular}{c|ccc} \hline
      had & HVP LO & HVP HO & HLbL\\
          & \multicolumn{3}{c}{$a_\mu [10^{-11}]$ (uncertainty)} \\\hline
      WP2020 & 6931 (40) & $-85.9 (0.7)$ & 92 (18) \\
      WP2025 & 7132 (61) & $-87.2 (1.3)$ & 115.5 (9.9) \\\hline
    \end{tabular}
\end{table}

Depending on which SM prediction one uses to compare with the direct measurement, dominated by the Fermilab measurement~\cite{Muong-2:2025xyk}, the conclusion is very different.
The $e^+e^-$ data-based prediction from WP2020 shows a discrepancy of more than 5 standard deviations with the measurement which could have been a first indication of new physics.
Unfortunately, with the new CMD3 measurement~\cite{PhysRevD.109.112002,PhysRevLett.132.231903}, the discrepancy among different measurements becomes so large that a new $e^+e^-$ data-based prediction becomes impossible. 
Therefore for WP2025, the lattice-based calculation, though with larger uncertainty, was used for the SM prediction which agrees with the measurement.

In Section~\ref{sec:review}, I shall show how the $e^+e^-$ data-based HVP calculation is performed and what the issues are with the $e^+e^-$ data.
The new analysis and the corresponding preliminary result from the BABAR Collaboration are then presented in Section~\ref{sec:babar}, followed by a summary in Section~\ref{sec:summary}.

\section{Review of the current $e^+e^-$ data-based HVP predictions} \label{sec:review}

The LO HVP contribution involves light quark loops which cannot be calculated at low energies from first principles.
Instead, based on the analyticity and unitarity argument, it can be calculated using $e^+e^-\to$ hadrons cross sections in terms of ratio $R(s)$ of the hadronic cross sections over the point like $e^+e^-\to \mu^+\mu^-$ cross section in the following dispersion relation:
\begin{equation}
a_\mu^\text{HVP LO}=\frac{\alpha^2}{3\pi}\int^\infty_{4m^2_\pi}ds\frac{K(s)}{s}R(s)\,,
\end{equation}
where $\alpha$ is the QED coupling constant and $K(s)$ is a QED kernel that has an energy dependence close to $1/s$ such that the low energy processes are highly weighted. 
Therefore, the precision of the $e^+e^-$ data-based HVP calculation is driven by that of the $e^+e^-$ to hadron cross-section measurements.
Note that the hadronic cross section has to be bare without vacuum polarization contribution to avoid double counting with the HO corrections, namely
$\sigma^0(s)[e^+e^-\to \mathrm{hadrons}(\gamma_\mathrm{FSR})]=\sigma(s)[e^+e^-\to \mathrm{hadrons}(\gamma_\mathrm{FSR})]\left(\alpha(0)/\alpha(s)\right)^2$.
The running QED coupling constant $\alpha(s)$ itself receives both leptonic and hadronic contributions as:
$\alpha(s)=\frac{\alpha(0)}{1-\Delta\alpha_\mathrm{lep}(s)-\Delta\alpha_\mathrm{had}(s)}$,
with
$\Delta\alpha_\mathrm{had}(s)=-\frac{\alpha(0)}{3\pi}\int^\infty_{4m^2_\pi}ds^\prime\frac{R(s)}{s^\prime(s^\prime-s-i\epsilon)}$.
Therefore, the HVP contribution is also relevant for the running QED coupling, except that the QED kernel has a different energy dependence, so that the higher energy part receives relatively larger weights.
Improving the HVP precision is also important here since it is currently a limiting factor for stringent SM consistency tests.

\begin{figure}[hbtp]
    \centering
    \includegraphics[width=0.4\linewidth]{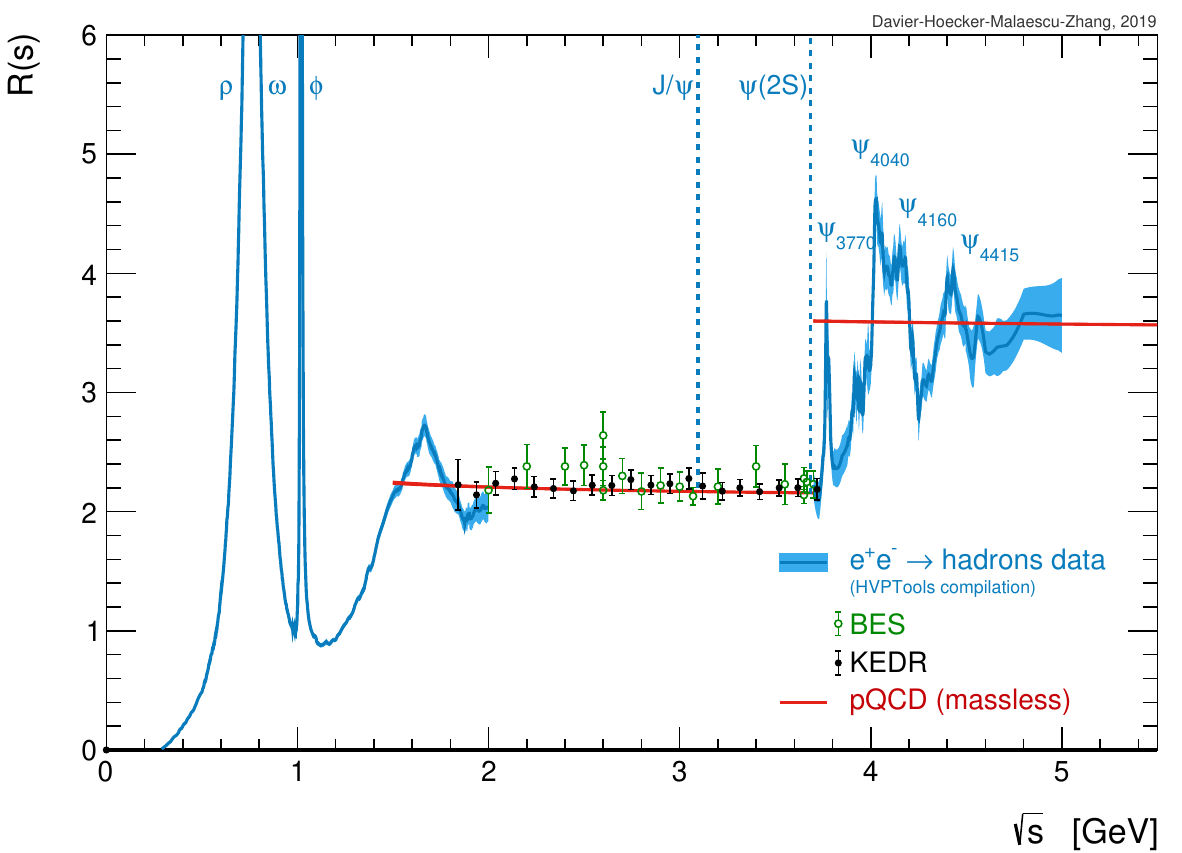}
    \includegraphics[width=0.59\linewidth]{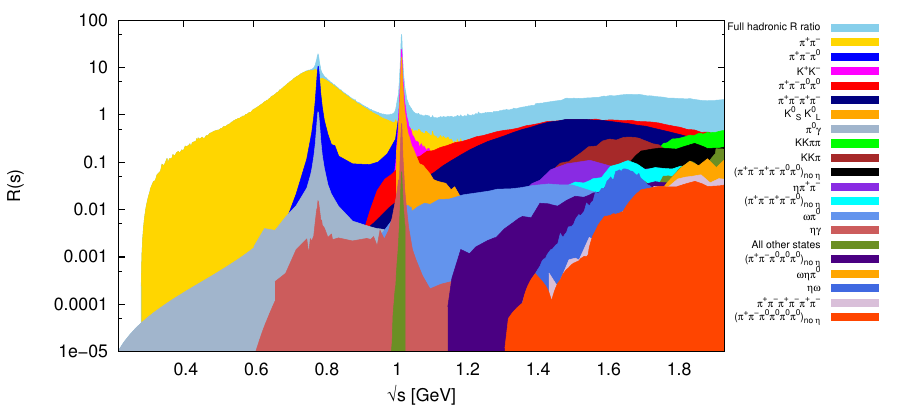}
    \caption{\small Left: the total hadronic $e^+e^-$ annihilation rate $R(s)$ as a function of center-of-mass energy $\sqrt{s}$. 
    Inclusive measurements from BES and KEDR are shown as data points, while the narrow blue bands correspond to the sum of exclusive channels, obtained using HVPTools~\cite{Davier:2010rnx}. Also shown for the purpose of illustration is the prediction from massless perturbative QCD (solid red line). Figure taken from Ref.~\cite{Davier:2019can}.
    Right: Contributions to the total hadronic $R$-ratio from different final states at low energies. The full $R$-ratio is shown in light blue. Each final state is included as a new layer on top in decreasing order of the size of its contribution to $a_\mu^\text{HVP LO}$. Figure taken from Ref.~\cite{Keshavarzi:2018mgv}.}
    \label{fig:rs}
\end{figure}

The energy dependence of the $R(s)$ ratio below 5\,GeV is shown in Figure~\ref{fig:rs} (left).
Between threshold and 1.8\,GeV, there are over 30 exclusive contributing channels.
The missing unmeasured channels are now negligible~\cite{Davier:2019can}.
Between 1.8 and 3.7\,GeV, the measurements are in good agreement with perturbative QCD (pQCD) prediction~\footnote{Some tension with the recent BESIII measurement~\cite{BESIII:2021wib} is observed and needs to be better understood.}.
Between 3.7 and 5\,GeV, there are rich structures, the combined data are directly used.
Above 5\,GeV, the pQCD prediction is again used.
The main exclusive channels below about 1.8\,GeV are shown explicitly in logarithmic $y$ scale in Figure~\ref{fig:rs} (right).
The dominant process is the two-pion channel, which gives about 73\% to the LO HVP contribution.
It also gives the largest contribution to the uncertainty budget.
Therefore, let us concentrate on the two-pion channel for the moment.

A compilation of all the measurements of the two-pion channel is presented in Table~\ref{tab:2pi}.
The top block concerns the measurements performed using the beam scan method, while the bottom one lists the measurements performed with the radiative return method.
A few comments are in order.
The old measurements before 2000 were no longer competitive in precision.
Some of the measurements, in particular the old one, need corrections when the measurements were not the bare cross sections.
Only few measurements were performed in ratio $\pi\pi/\mu\mu$ allowing the cancellation of some of the systematic uncertainties and avoiding
dependence on $e^+e^-$ luminosity measurement.
Even fewer measurements were performed in blinded ways.
New measurements have increasing precision but show, however, already large discrepancy 
between BABAR (2009)~\cite{BaBar:2012bdw,BaBar:2009wpw} and KLOE (2008-2012)~\cite{KLOE:2008fmq,KLOE:2010qei,KLOE:2012anl} for WP2020.
Since WP2020, three new measurements, SND (2020)~\cite{SND:2020nwa}, CMD3 (2023) and BABAR (2025), are available.
The new CMD3 results are larger than all other measurements, including CMD2.
The new BABAR (2025) measurement is essentially independent of BABAR (2009), see below.

\begin{table}[htbp]
    \centering
    \caption{\small A compilation of all measurements of the two-pion channel showing the number of measurement points, energy coverage and the corresponding uncertainties.}
    \label{tab:2pi}
    \vspace{2mm}
\begin{tabular}{l|c|c|c|c}
\hline
Experiment & $N_{\rm data}$ & Energy range [GeV] & $\delta$(stat.) [\%] & $\delta$(syst.) [\%] \\ \hline
\multicolumn{5}{c}{Beam energy scan}\\ \hline
DM1 (1978) & $16$ & $0.483-1.096$ & $6.6-40$ & $2.2$ \\
TOF (1981) & $4$ & $0.400-0.460$ & $14-20$ & $5$ \\
OLYA (1979, 1985) & $2+77$ & $0.400-1.397$ & $2.3-35$ & $4$ \\
CMD (1985) & $24$ & $0.360-0.820$ & $4.1-10.8$ & $2$ \\
DM2 (1989) & $17$ & $1.350-2.215$ & $17.6-100$ & $12$ \\ 
CMD2 (2003) & $43$ & $0.611-0.962$ & $1.8-14.1$ & $0.6$ \\
SND (2006) & $45$ & $0.390-0.970$ & $0.5-2.1$ & $1.2-3.8$ \\
CMD2$_{\rm low}$ (2006) & $10$ & $0.370-0.520$ & $4.5-7$ & $0.7$ \\
CMD2$_{\rm rho}$ (2006) & $29$ & $0.600-0.970$ & $0.5-4.1$ & $0.8$ \\
CMD2$_{\rm high}$ (2006) & $36$ & $0.980-1.380$ & $4.5-18.4$ & $1.2-4.2$ \\
SND (2020) & 36 & $0.525-0.883$ & $0.4-6.0$ & $0.8-1.2$ \\
CMD3 (2023) & 151 & $0.32-1.2$ & $0.2-6.8$ & $0.7-2.0$\\ \hline
\multicolumn{5}{c}{ISR (initial state radiation) photon / radiative return}\\ \hline
KLOE (2008) & 60 & $0.592-0.975$ & $0.3-1.2$ & 0.9 \\
BABAR (2009) & 337 & $0.3-3.0$ & $0.9-10.5$ {\tiny ($<1.4$\,GeV)} & $0.5-1.4$ {\tiny ($<1.4$\,GeV)} \\
KLOE (2010) & 75 & $0.316-0.922$ & $0.6-17.6$ & $1.3-10$ \\
KLOE (2012) & 60 & $0.592-0.975$ & $1.8-2.6$ & 0.9 \\
BESIII (2015) & 60 & $0.6-0.9$ & $2.0-4.3$ & 0.9 \\
CLEO (2017) & 35 & $0.3-1.0$ & $2-20$ & 1.5 \\
BABAR (2025) & 311 & $0.28-1.4$ & down to 0.5 & down to 0.3 \\
\hline
\end{tabular}
\end{table}

\begin{figure}[hbtp]
    \centering
    \includegraphics[width=0.49\linewidth]{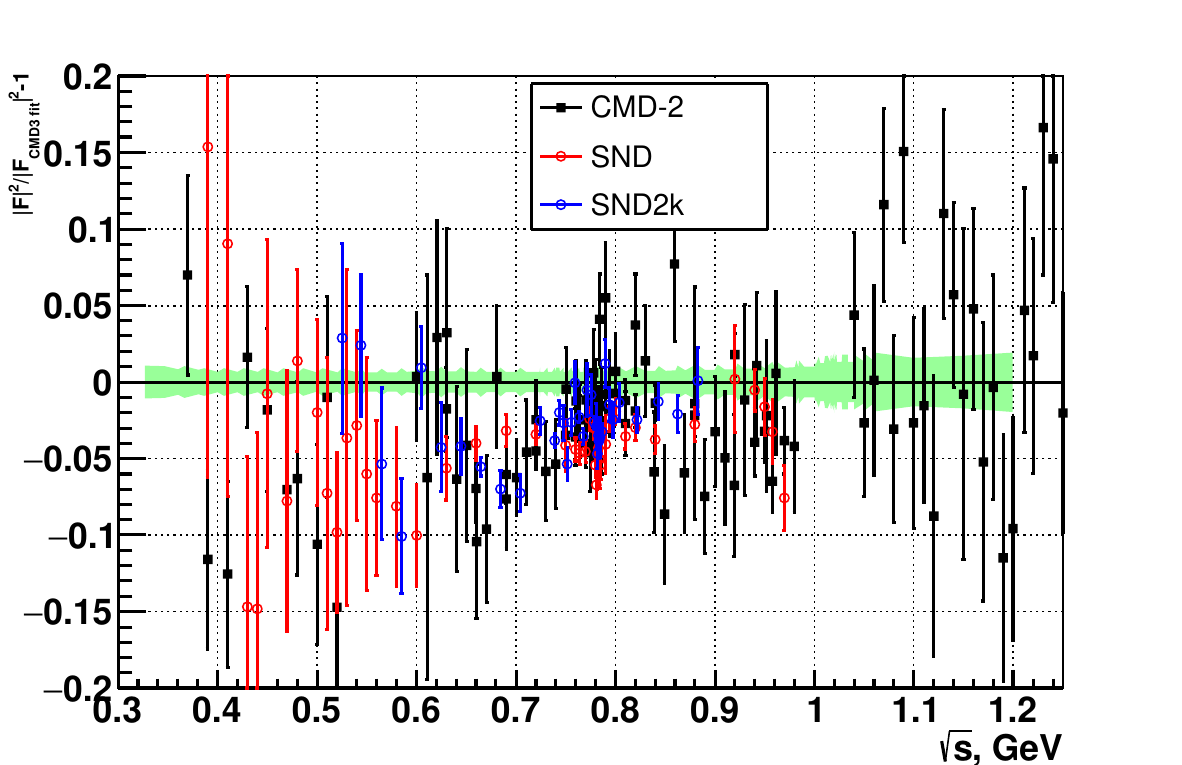}
    \includegraphics[width=0.49\linewidth]{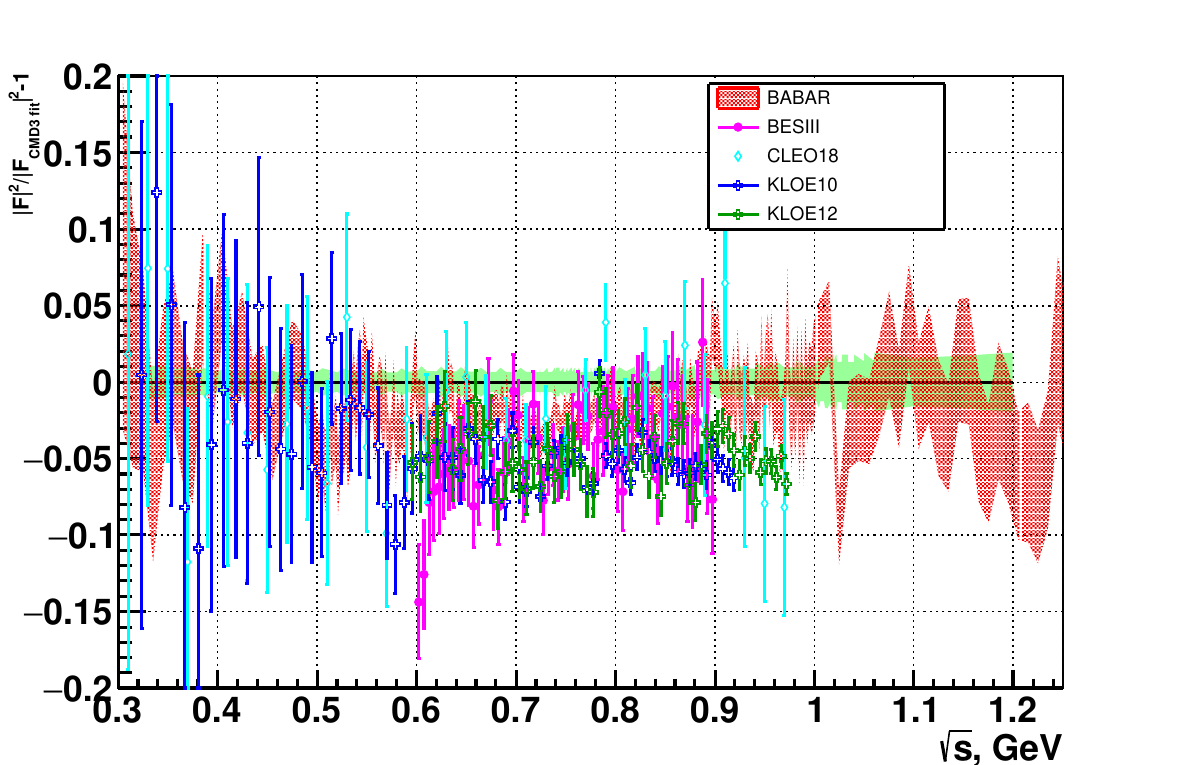}
    \caption{\small Relative differences in terms of pion form factors between CMD3 and previous scan-based measurements (left) and ISR photon-based measurements (right). The green band corresponds to the systematic uncertainty of the CMD-3 measurement. Figures taken from Ref.~\cite{CMD-3:2023alj}.}
    \label{fig:cmd3-en}
\end{figure}

In Figure~\ref{fig:cmd3-en}, the recent CMD3 measurement is compared on the left as a function of energy with the other energy scan measurements from CMD2 and SND at the same collider and on the right with the radiative return measurements.
There is a fairly good agreement with CMD2 above 1\,GeV and with BABAR above 0.9\,GeV.
The agreement with BABAR and KLOE at low energy is also fairly good.
Elsewhere, CMD3 tends to be higher than all other measurements.
The measurements density versus energy and energy coverage are very different between different experiments.

\begin{figure}[hbtp]
    \centering
    \includegraphics[width=0.49\linewidth]{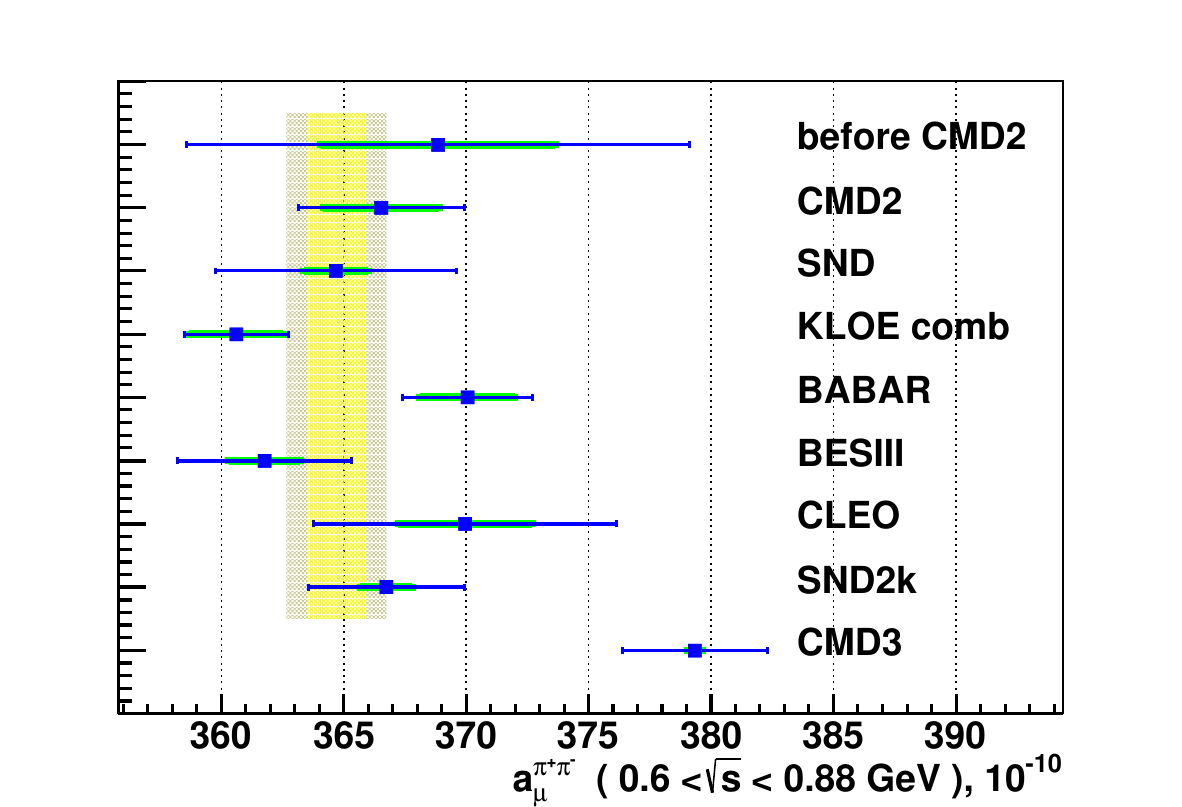}
    \includegraphics[width=0.4\linewidth]{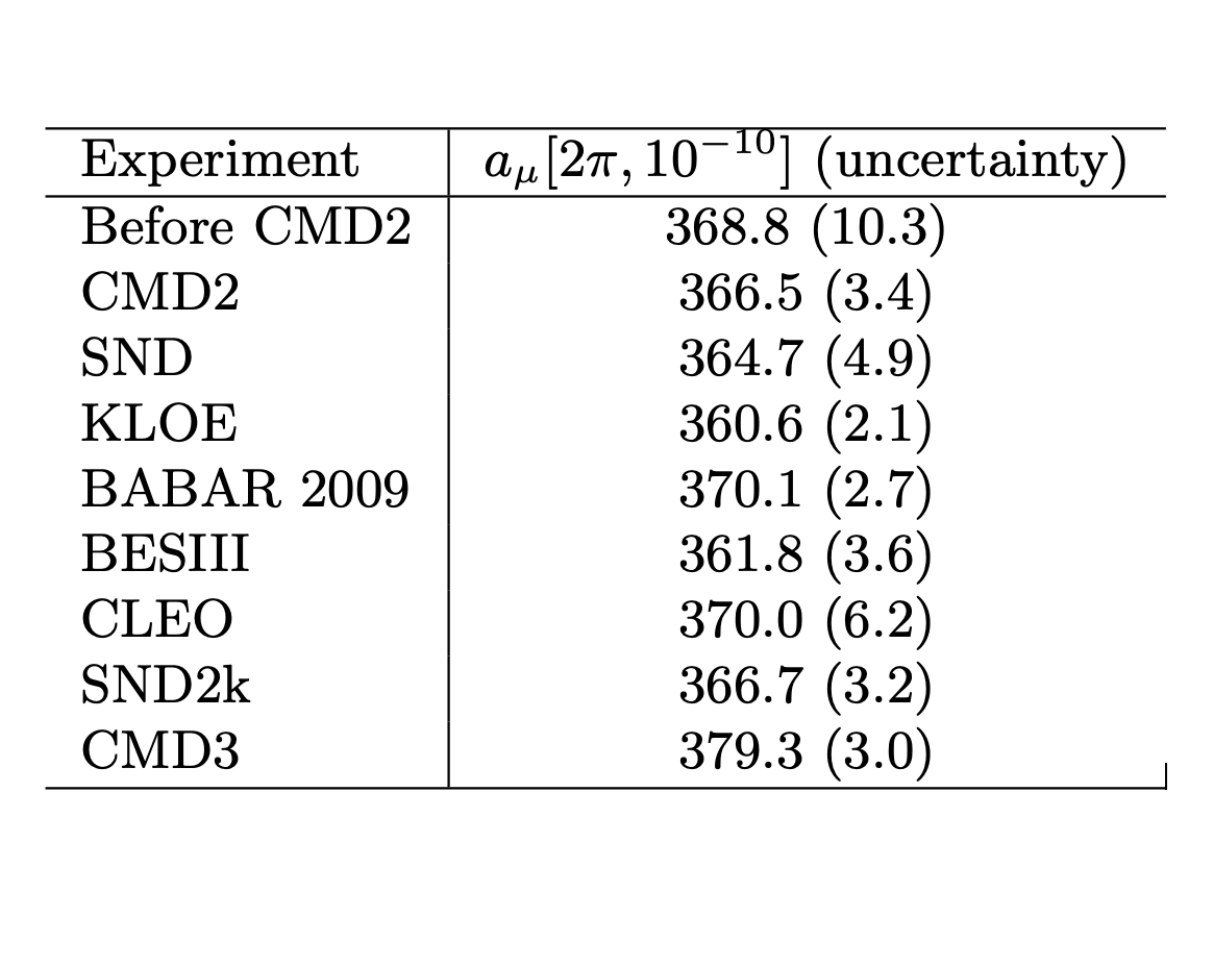}
    \caption{\small The $\pi\pi(\gamma)$ contribution to the LO HVP
from the energy range $0.6 <\sqrt{s} < 0.88$\,GeV obtained
from the CMD3 data and the results of the other experiments. Figure (Table) taken (adapted) from Ref.~\cite{CMD-3:2023alj}.}
    \label{fig:amu_comp}
\end{figure}

A different way to compare different measurements is to look at their contributions to the LO HVP in the common energy range between 0.6 and 
0.88\,GeV both graphically and numerically as shown in Figure~\ref{fig:amu_comp}.
The discrepancy is clearly seen between KLOE and BABAR and 
becomes worse between KLOE and CMD3.
This prevents a new combination for the $e^+e^-$ data-based HVP prediction for WP2025.


In addition to the two-pion channel, there are many other channels contributing to the HVP calculations.
The BABAR experiment have measured essentially all of them. 
Since WP2020, several new measurements have been performed in either the existing channels or new ones~\cite{BABAR:2021cde,BaBar:2021rki,BaBar:2021gyu,BaBar:2022ahi}, such that the relative contributions to $a_\mu$ from the unmeasured channels have been reduced from 0.9\% in 2003 to a negligible level nowadays.

The tension among the different measurements in the dominant two-pion channel is not the only one.
There are in fact tensions also in other channels.
For instance, in the three-point channel, the new measurement from Belle II shows tension with BABAR, SND and CMD2, in the dominant omega resonance region. This channel contributes by about 10\% to the HVP calculation and is the 2nd largest channel next to the two-pion channel.
The situation is similar in the $KK$ channel, which contributes by about 3.3\% to the HVP calculation and is the 3rd most important channel.  
There are again tensions between different measurements.

\section{New BABAR analysis and result} \label{sec:babar}

Given the current confusing situation, it is extremely important to perform precise and independent measurements. 
Here, we present for the first time a brand new measurement of the two-pion cross section from the BABAR experiment.
The new analysis uses the full data sample corresponding to an integrated luminosity of 460\,fb$^{-1}$ which is about a factor of 2 larger than the previous one in 2009.
The new analysis also exploits different angular distributions to separate the $\pi\pi$ and $\mu\mu$ final states, in contrast with particle identification used in the previous analysis.  

Using the full data sample, BABAR has recently also carried out a unique HO radiation study with up to three hard radiation photons at small and large angles in the initial and final states in the two-pion and dimuon channels~\cite{BaBar:2023xiy}. Two key observations are: (1) the fraction of events with next-to-next-leading radiation (i.e.\ three hard photons) in data is observed to be about 3.5\%, which is absent in the current next-to-leading order event generator \textsc{Phokhara}; (2) \textsc{Phokhara} predicts about 25\% too high rate for hard next-to-leading order ISR photons at small angles. The consequences of the observations were studied in Refs.~\cite{BaBar:2023xiy,Davier:2023fpl} and it turned out the 2009 BABAR measurement is unaffected given that the event selections were inclusive for the high-order radiations and the efficiency corrections were evaluated with data. The other measurements relying heavily on \textsc{Phokhara} predictions may be affected and need to be checked by the relevant experiments.

The key analysis steps of the new BABAR measurement include: (1) the background suppression using similar selections as the 2009 analysis; (2) evaluate data-MC differences and their uncertainties affecting the angular (template) and mass distributions in terms of trigger, tracking and event selections; and (3) perform kinematic (template) fits to separate the two-pion and dimuon final states. The analysis is performed in blinded ways with three different blinding factors for the trigger and tracking efficiency corrections as well as the normalization of the fitted two-pion/dimuon mass spectra for each of the two-pion and dimuon processes. 

\begin{figure}[hbtp]
    \centering
    \includegraphics[width=0.49\linewidth]{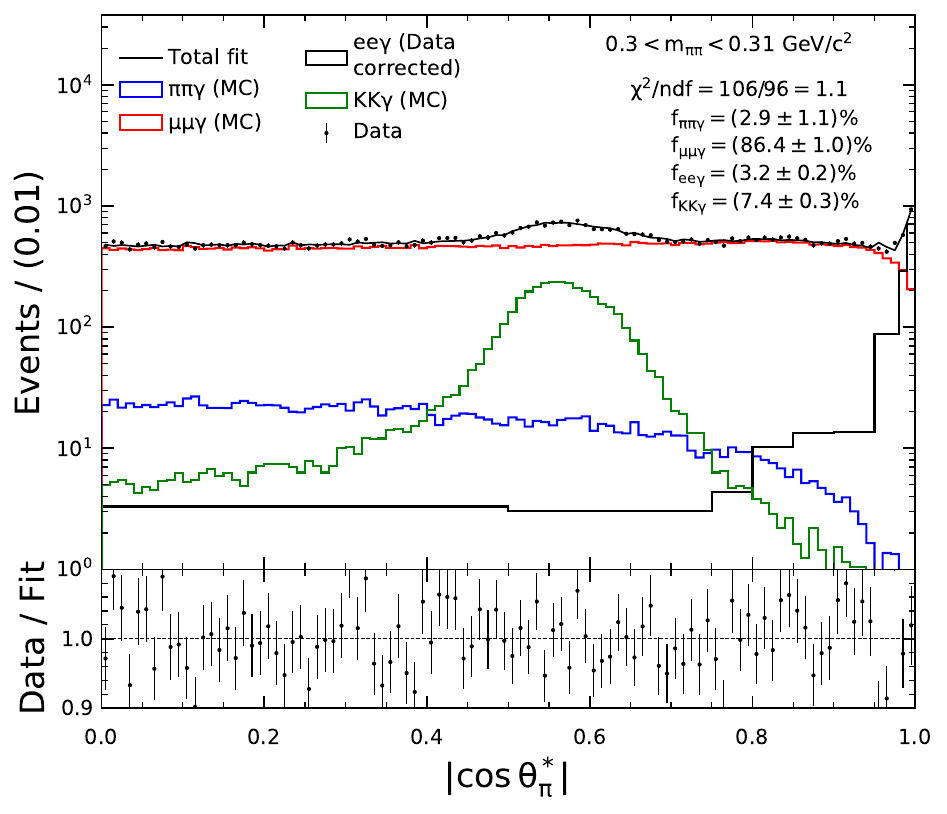}
    \includegraphics[width=0.49\linewidth]{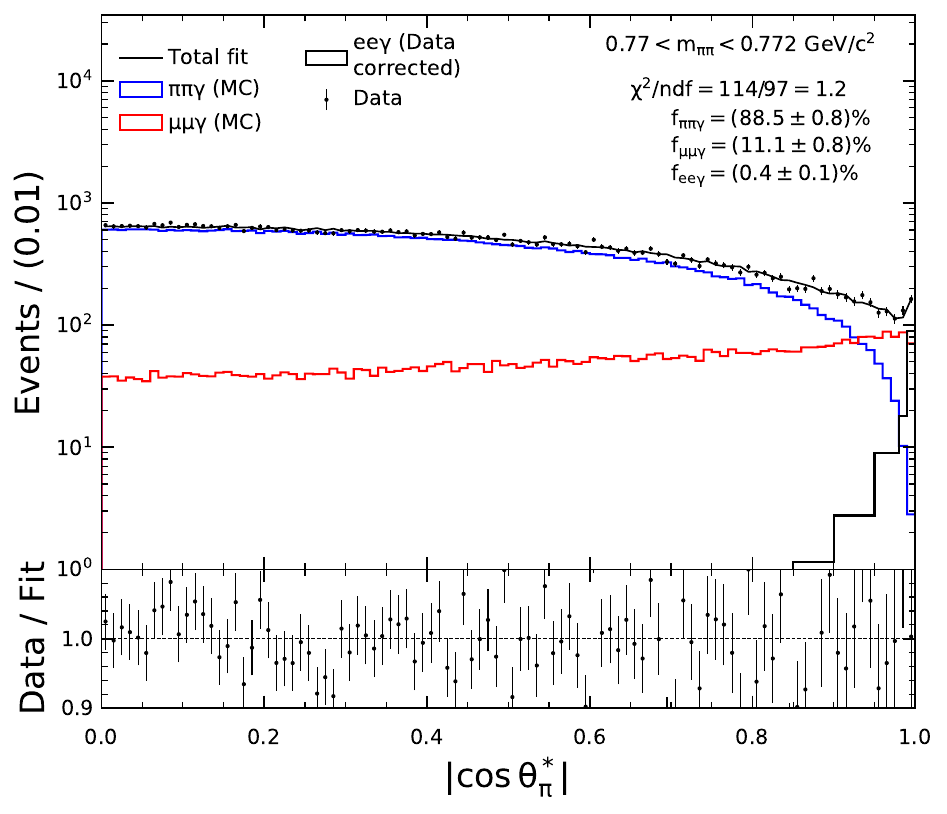}
    \caption{\small Illustration of the kinematic (template) fit for the signals ($\pi\pi$ or $\mu\mu$) and background ($KK$ and $ee\gamma$) separation for two selected mass intervals at $0.30<m_{\pi\pi}<0.31$\,GeV/c$^2$ (left) and $0.770<m_{\pi\pi}<0.772$\,GeV/c$^2$ (right).}
    \label{fig:fit}
\end{figure}

The fit strategy is optimized to minimize the uncertainties of the $\mu\mu$ and $\pi\pi$ processes. The fit procedure for a given mass interval (typically 2\,MeV/c$^2$ between 0.5 and 1\,GeV/c$^2$ and 10\,MeV/c$^2$ elsewhere) is the following: (1) an initial fit in the large $|\cos\theta^*|$ tail 0.9--1 is performed to fix the normalization of the $ee\gamma$ background using a mass dependent $ee\gamma$ template obtained from data; (2) a new fit is performed in the region $|\cos\theta^*|<0.9$ to separate $\mu\mu$ and $\pi\pi$ processes (also $KK$ for low $m_{\pi\pi}$) after having subtracted the small residual $ee\gamma$ contribution (the other small background from all other processes has already been subtracted from the data distribution); and (3) the fitted $\mu\mu$ and $\pi\pi$ event yields are then extrapolated to the full $|\cos\theta^*|$ 0--1 range. The $\mu\mu$ and $\pi\pi$ templates used in the fits and extrapolation are obtained from the \textsc{Phokhara} simulation but corrected for any data/MC differences. The data distribution and the different signal and background components for two mass intervals are shown in Figure~\ref{fig:fit} to illustrate the fit.

\begin{figure}[hbtp]
    \centering
    \includegraphics[width=0.49\linewidth]{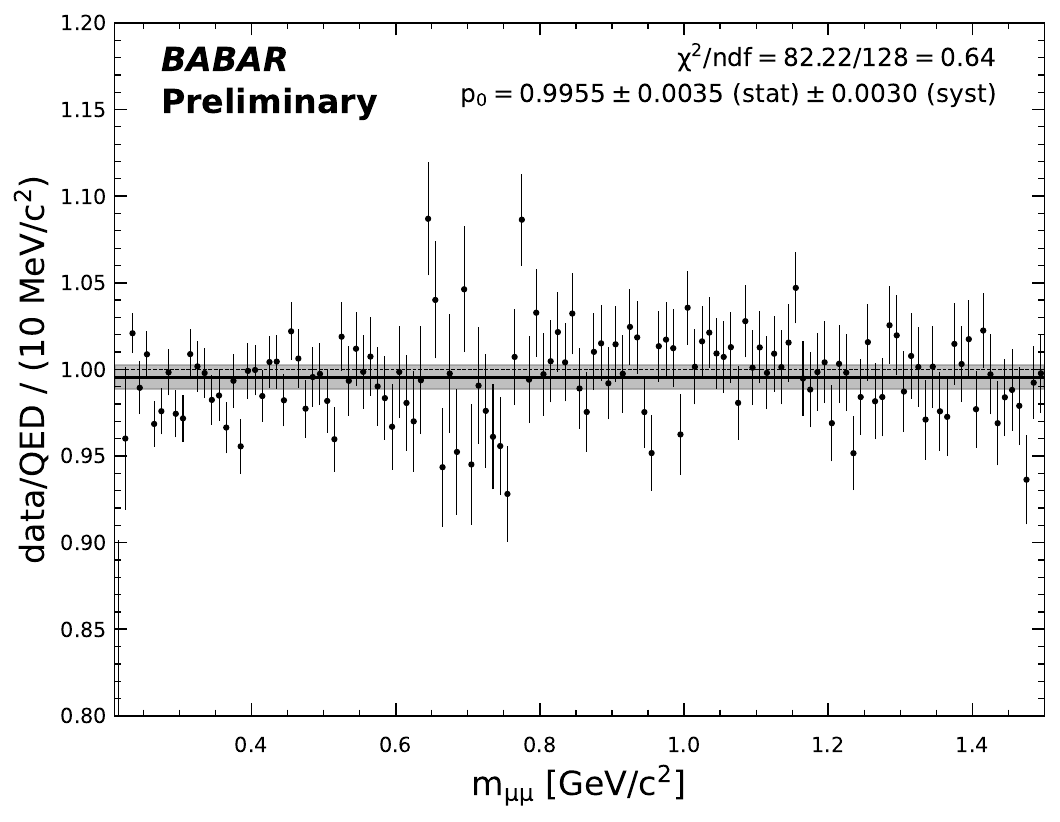}
    \caption{\small The ratio of the unblinded $\mu\mu$ mass
spectrum over the corrected \textsc{Phokhara} (QED)
prediction. The full horizontal line represents a
constant fit to the points. The gray band shows the total systematic uncertainty
(including contributions from the ISR photon
data/MC efficiency and the $e^+e^-$ luminosity).}
    \label{fig:test-QED}
\end{figure}

When all the studies have been finalized, the dimuon channel was first unblinded to compare the dimuon event yields as a function of mass with the QED prediction. The result is shown in Figure~\ref{fig:test-QED} and the corresponding ratio is $R_{\mu\mu}=0.9955\pm 0.0035_\mathrm{stat}\pm 0.0030_\mathrm{syst}\pm 0.0033_\text{$\gamma$ ISR}\pm 0.0033_\text{lumi $ee$}$. This is a non-trivial test as each data/MC correction affects not only the $\pi\pi/\mu\mu$ separation but also the $\mu\mu$ mass spectrum. The unblinded $\mu\mu$ mass spectrum is then unfolded to get the $\mu\mu$ spectrum as a function of $\sqrt{s^\prime}$, where $s^\prime=s(1-2E^*_\gamma/\sqrt{s})$ with $s$ being the $e^+e^-$ center-of-mass energy squared. Using the unfolded $\mu\mu$ energy spectrum together with the acceptance and the bare cross section of the $\mu\mu$ process, the effective luminosity of the ISR process as a function of $\sqrt{s^\prime}$ is derived. 

\begin{figure}[hbtp]
    \centering
    \includegraphics[width=0.49\linewidth]{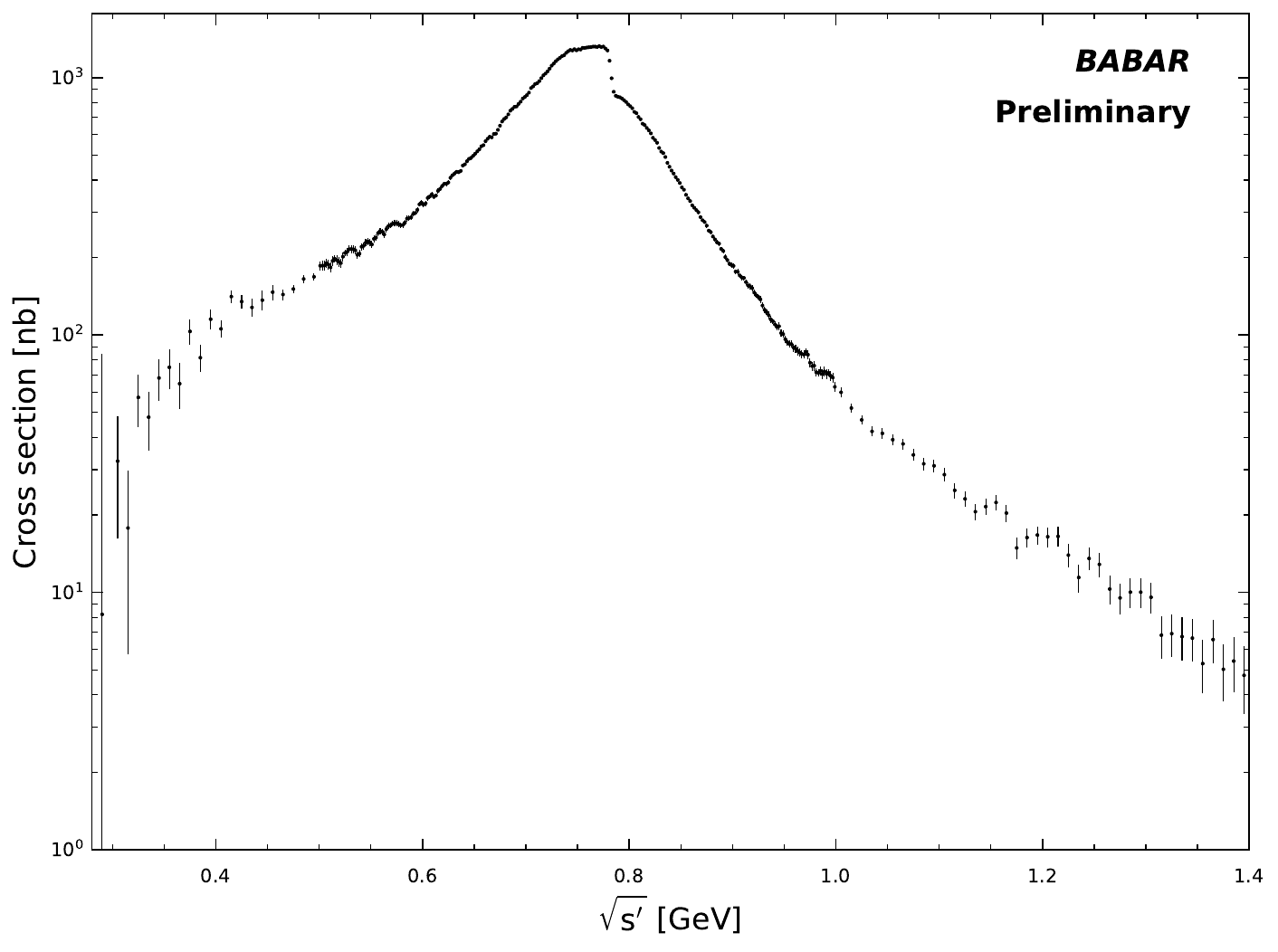}
    \includegraphics[width=0.49\linewidth]{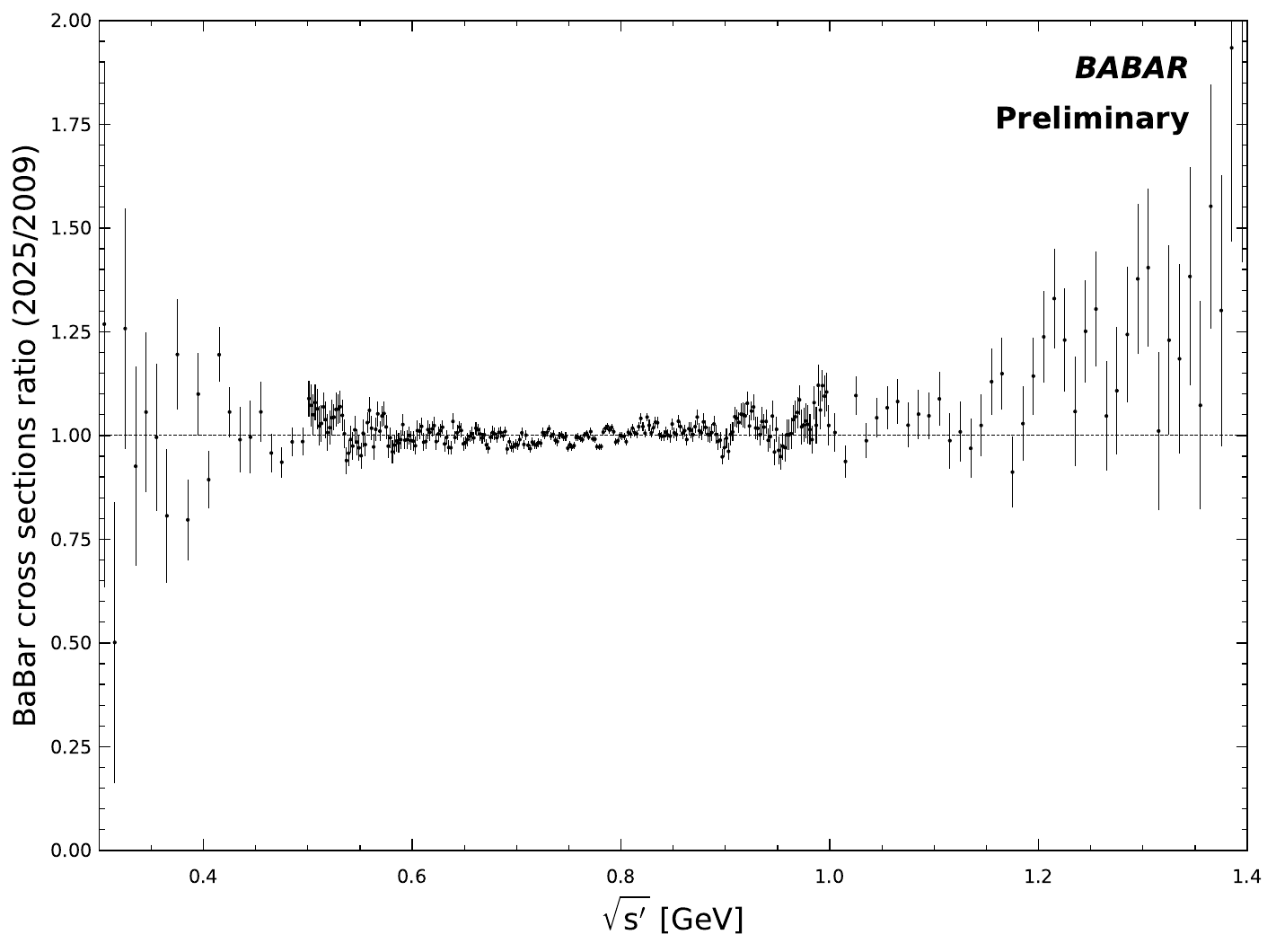}
    \caption{\small The measured preliminary cross section for $e^+e^-\to \pi^+\pi^-(\gamma)$ over the full energy range (left) and the ratio of the new measurement and the previous one in 2009, from 0.3 to 1.4\,GeV (right).}
    \label{fig:xs-2pi}
\end{figure}

With the successful QED test for the $\mu\mu$ channel, the $\pi\pi$ channel was then unblinded. The $\pi\pi$ cross section is obtained using the unfolded $\pi\pi$ energy spectrum, its acceptance correction and the derived ISR effective luminosity. The result is shown in Figure~\ref{fig:xs-2pi} (left) and compared with the corresponding one from 2009 (right). The new measurement, which is more precise in the rho peak region, is in excellent agreement with the previous one, which has better precision in the low and high energy regions. The different precisions in different energy regions reflect the complementarity nature of the two methods. In term of its contribution to $a_\mu$, the preliminary 2025-2009 combination gives $a_\mu [2\pi, 1.8\,\mathrm{GeV}]=514.4 (0.49\%)\times 10^{-10}$ providing the most precise measurement in this channel.

\section{Summary} \label{sec:summary}

\begin{figure}[hbtp]
    \centering
    \includegraphics[width=0.5\linewidth]{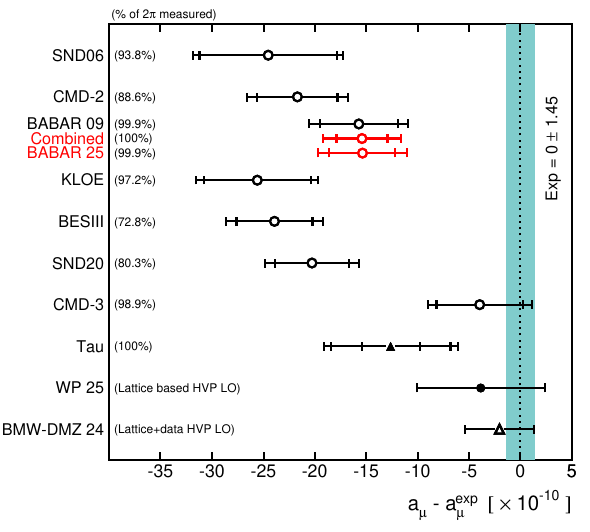}
    \caption{\small Compilation of $a_\mu$ predictions subtracted by the central value of the experimental world average~\cite{Muong-2:2025xyk}. The predictions are computed from the individual $\pi^+\pi^-$ contributions between threshold and 1.8\,GeV, complemented by common non-$\pi^+\pi^-$ contributions. The error bars indicate the $\pi^+\pi^-$ and total uncertainties, respectively. The intermediate error bar for tau corresponds to the inflated uncertainty of isospin breaking corrections. The percentage represents the fraction of $a_\mu[\pi^+\pi^-, \text{threshold–1.8\,GeV}]$ used from a given experiment. The tau and lattice-based WP2025 results~\cite{Aliberti:2025beg} are shown with filled triangle and dot, respectively. The combined lattice and data result from BMW-DMZ~\cite{Boccaletti:2024guq} is also shown with open triangle.}
    \label{fig:amu-summary}
\end{figure}

The $e^+e^-$ data-based LO HVP calculation is mainly contributed by the two-pion channel. There is, however, large discrepancies among different measurements which prevent a new data-driven prediction for WP2025. Therefore, the new SM WP2025 prediction is based on LO HVP lattice calculations.
The new BABAR measurement in the two-pion channel performed in fully blinded ways confirms the previous measurement in 2009. The preliminary BABAR 2009-2025 combination provides the most precise measurement in the two-pion channel. The current status of the comparison between either the data-driven predictions or lattice-based predictions and the direct measurement is summarized in Figure~\ref{fig:amu-summary}. An alternative data-driven prediction~\cite{Aliberti:2025beg} using tau data proposed in Ref.~\cite{Alemany:1997tn} is also shown.

We need better event generators and more precise and independent measurements from other experiments to clarify the situation. The difference between the $e^+e^-$ data-based HVP calculation and the lattice-based one also needs to be understood before a firm conclusion whether there is or not any new physics in the muon $g-2$ observable can be drawn.

\printbibliography

@article{Alemany:1997tn,
    author = "Alemany, Ricard and Davier, Michel and Hocker, Andreas",
    title = "{Improved determination of the hadronic contribution to the muon ($g-2$) and to $\alpha(M^2_Z)$ using new data from hadronic $\tau$ decays}",
    eprint = "hep-ph/9703220",
    archivePrefix = "arXiv",
    reportNumber = "LAL-97-02",
    doi = "10.1007/s100520050127",
    journal = "Eur. Phys. J. C",
    volume = "2",
    pages = "123--135",
    year = "1998"
}

@article{Davier:2023fpl,
    author = "Davier, Michel and Hoecker, Andreas and Lutz, Anne-Marie and Malaescu, Bogdan and Zhang, Zhiqing",
    title = "{Tensions in $e^+e^-\rightarrow \pi ^+\pi ^-(\gamma )$ measurements: the new landscape of data-driven hadronic vacuum polarization predictions for the muon $g - 2$}",
    eprint = "2312.02053",
    archivePrefix = "arXiv",
    primaryClass = "hep-ph",
    doi = "10.1140/epjc/s10052-024-12964-7",
    journal = "Eur. Phys. J. C",
    volume = "84",
    number = "7",
    pages = "721",
    year = "2024"
}

@article{BaBar:2023xiy,
    author = "Lees, J. P. and others",
    collaboration = "BaBar",
    title = "{Measurement of additional radiation in the initial-state-radiation processes $e^+e^-\to\mu^+\mu^-\gamma$ and $e^+e^-\to\pi^+\pi^-\gamma$ at BABAR}",
    eprint = "2308.05233",
    archivePrefix = "arXiv",
    primaryClass = "hep-ex",
    reportNumber = "BABAR-PUB-23/005, SLAC-PUB-17736",
    doi = "10.1103/PhysRevD.108.L111103",
    journal = "Phys. Rev. D",
    volume = "108",
    number = "11",
    pages = "L111103",
    year = "2023"
}

@article{SND:2020nwa,
    author = "Achasov, M. N. and others",
    collaboration = "SND",
    title = "{Measurement of the $e^+e^- \to\pi^+\pi^- $ process cross section with the SND detector at the VEPP-2000 collider in the energy region $0.525<\sqrt{s}<0.883$ GeV}",
    eprint = "2004.00263",
    archivePrefix = "arXiv",
    primaryClass = "hep-ex",
    doi = "10.1007/JHEP01(2021)113",
    journal = "JHEP",
    volume = "01",
    pages = "113",
    year = "2021"
}

@article{KLOE:2012anl,
    author = "Babusci, D. and others",
    collaboration = "KLOE",
    title = "{Precision measurement of $\sigma(e^+e^-\rightarrow \pi^+\pi^-\gamma)/ \sigma(e^+e^-\rightarrow \mu^+\mu^-\gamma)$ and determination of the $\pi^+\pi^-$ contribution to the muon anomaly with the KLOE detector}",
    eprint = "1212.4524",
    archivePrefix = "arXiv",
    primaryClass = "hep-ex",
    doi = "10.1016/j.physletb.2013.02.029",
    journal = "Phys. Lett. B",
    volume = "720",
    pages = "336--343",
    year = "2013"
}

@article{KLOE:2010qei,
    author = "Ambrosino, F. and others",
    collaboration = "KLOE",
    title = "{Measurement of $\sigma(e^+ e^- \to \pi^+ \pi^-)$ from threshold to 0.85\,GeV$^2$ using Initial State Radiation with the KLOE detector}",
    eprint = "1006.5313",
    archivePrefix = "arXiv",
    primaryClass = "hep-ex",
    doi = "10.1016/j.physletb.2011.04.055",
    journal = "Phys. Lett. B",
    volume = "700",
    pages = "102--110",
    year = "2011"
}

@article{KLOE:2008fmq,
    author = "Ambrosino, F. and others",
    collaboration = "KLOE",
    title = "{Measurement of $\sigma(e^+ e^- \to \pi^+ \pi^- \gamma(\gamma)$ and the dipion contribution to the muon anomaly with the KLOE detector}",
    eprint = "0809.3950",
    archivePrefix = "arXiv",
    primaryClass = "hep-ex",
    doi = "10.1016/j.physletb.2008.10.060",
    journal = "Phys. Lett. B",
    volume = "670",
    pages = "285--291",
    year = "2009"
}

@article{BaBar:2009wpw,
    author = "Aubert, Bernard and others",
    collaboration = "BaBar",
    title = "{Precise measurement of the $e^+ e^- \to \pi^+ \pi^- (\gamma)$ cross section with the Initial State Radiation method at BABAR}",
    eprint = "0908.3589",
    archivePrefix = "arXiv",
    primaryClass = "hep-ex",
    reportNumber = "SLAC-PUB-13754, BABAR-PUB-09-027",
    doi = "10.1103/PhysRevLett.103.231801",
    journal = "Phys. Rev. Lett.",
    volume = "103",
    pages = "231801",
    year = "2009"
}

@article{BaBar:2012bdw,
    author = "Lees, J. P. and others",
    collaboration = "BaBar",
    title = "{Precise Measurement of the $e^+ e^- \to \pi^+\pi^- (\gamma)$ Cross Section with the Initial-State Radiation Method at BABAR}",
    eprint = "1205.2228",
    archivePrefix = "arXiv",
    primaryClass = "hep-ex",
    reportNumber = "BABAR-PUB-12-003",
    doi = "10.1103/PhysRevD.86.032013",
    journal = "Phys. Rev. D",
    volume = "86",
    pages = "032013",
    year = "2012"
}

@article{Boccaletti:2024guq,
    author = "Boccaletti, A. and others",
    title = "{High precision calculation of the hadronic vacuum polarisation contribution to the muon anomaly}",
    eprint = "2407.10913",
    archivePrefix = "arXiv",
    primaryClass = "hep-lat",
    month = "7",
    year = "2024"
}

@article{BaBar:2022ahi,
    author = "Lees, J. P. and others",
    collaboration = "BaBar",
    title = "{Study of the reactions $e^+e^-\to K^+K^-\pi^0\pi^0\pi^0$, $e^+e^-\to K_S^0K^\pm\pi^\mp\pi^0\pi^0$, and $e^+e^-\to K_S^0 K^\pm\pi^\mp\pi^+\pi^-$ at center-of-mass energies from threshold to 4.5\,GeV using initial-state radiation}",
    eprint = "2207.10340",
    archivePrefix = "arXiv",
    primaryClass = "hep-ex",
    reportNumber = "SLAC-PUB-17694",
    doi = "10.1103/PhysRevD.107.072001",
    journal = "Phys. Rev. D",
    volume = "107",
    number = "7",
    pages = "072001",
    year = "2023"
}

@article{BaBar:2021gyu,
    author = "Lees, J. P. and others",
    collaboration = "BaBar",
    title = "{Study of the reactions $e^+e^-\to\pi^+\pi^-\pi^0\pi^0\pi^0\pi^0$ and $\pi^+\pi^-\pi^0\pi^0\pi^0\eta$ at center-of-mass energies from threshold to 4.5 GeV using initial-state radiation}",
    eprint = "2110.00823",
    archivePrefix = "arXiv",
    primaryClass = "hep-ex",
    reportNumber = "SLAC-PUB-17619, BABAR-PUB-21/005",
    doi = "10.1103/PhysRevD.104.112004",
    journal = "Phys. Rev. D",
    volume = "104",
    number = "11",
    pages = "112004",
    year = "2021"
}

@article{BaBar:2021rki,
    author = "Lees, J. P. and others",
    collaboration = "BaBar",
    title = "{Study of the reactions $e^+e^-\to2(\pi^+\pi^-)\pi^0\pi^0\pi^0$ and $e^+e^-\to2(\pi^+\pi^-)\pi^0\pi^0\eta$ at center-of-mass energies from threshold to 4.5 GeV using initial-state radiation}",
    eprint = "2102.01314",
    archivePrefix = "arXiv",
    primaryClass = "hep-ex",
    reportNumber = "SLAC-PUB-17587, BaBar-PUB-20004",
    doi = "10.1103/PhysRevD.103.092001",
    journal = "Phys. Rev. D",
    volume = "103",
    number = "9",
    pages = "092001",
    year = "2021"
}

@article{BABAR:2021cde,
    author = "Lees, J. P. and others",
    collaboration = "BABAR, BaBar",
    title = "{Study of the process $e^+e^-\to \pi^+\pi^-\pi^0$ using initial state radiation with BABAR}",
    eprint = "2110.00520",
    archivePrefix = "arXiv",
    primaryClass = "hep-ex",
    reportNumber = "BABAR-PUB-21/004, SLAC-PUB-17620",
    doi = "10.1103/PhysRevD.104.112003",
    journal = "Phys. Rev. D",
    volume = "104",
    number = "11",
    pages = "112003",
    year = "2021"
}

@article{CMD-3:2023alj,
    author = "Ignatov, F. V. and others",
    collaboration = "CMD-3",
    title = "{Measurement of the e+e-{\textrightarrow}{\ensuremath{\pi}}+{\ensuremath{\pi}}- cross section from threshold to 1.2~GeV with the CMD-3 detector}",
    eprint = "2302.08834",
    archivePrefix = "arXiv",
    primaryClass = "hep-ex",
    doi = "10.1103/PhysRevD.109.112002",
    journal = "Phys. Rev. D",
    volume = "109",
    number = "11",
    pages = "112002",
    year = "2024"
}

@article{Keshavarzi:2018mgv,
    author = "Keshavarzi, Alexander and Nomura, Daisuke and Teubner, Thomas",
    title = "{Muon $g-2$ and $\alpha(M_Z^2)$: a new data-based analysis}",
    eprint = "1802.02995",
    archivePrefix = "arXiv",
    primaryClass = "hep-ph",
    reportNumber = "LTH 1153, KEK-TH-2035, LTH-1153, YITP-18-09, LTH 1153; KEK-TH-2035; YITP-18-09",
    doi = "10.1103/PhysRevD.97.114025",
    journal = "Phys. Rev. D",
    volume = "97",
    number = "11",
    pages = "114025",
    year = "2018"
}

@article{BESIII:2021wib,
    author = "Ablikim, M. and others",
    collaboration = "BESIII",
    title = "{Measurement of the Cross Section for $e^{+}e^{-}\to$Hadrons at Energies from 2.2324 to 3.6710~GeV}",
    eprint = "2112.11728",
    archivePrefix = "arXiv",
    primaryClass = "hep-ex",
    doi = "10.1103/PhysRevLett.128.062004",
    journal = "Phys. Rev. Lett.",
    volume = "128",
    number = "6",
    pages = "062004",
    year = "2022"
}

@article{Davier:2019can,
    author = "Davier, M. and Hoecker, A. and Malaescu, B. and Zhang, Z.",
    title = "{A new evaluation of the hadronic vacuum polarisation contributions to the muon anomalous magnetic moment and to $\alpha(m_Z^2)$}",
    eprint = "1908.00921",
    archivePrefix = "arXiv",
    primaryClass = "hep-ph",
    doi = "10.1140/epjc/s10052-020-7792-2",
    journal = "Eur. Phys. J. C",
    volume = "80",
    number = "3",
    pages = "241",
    year = "2020",
    note = "[Erratum: Eur.Phys.J.C 80, 410 (2020)]"
}

@article{Davier:2010rnx,
    author = "Davier, M. and Hoecker, A. and Malaescu, B. and Yuan, C. Z. and Zhang, Z.",
    title = "{Reevaluation of the hadronic contribution to the muon magnetic anomaly using new e+ e- ---{\ensuremath{>}} pi+ pi- cross section data from BABAR}",
    eprint = "0908.4300",
    archivePrefix = "arXiv",
    primaryClass = "hep-ph",
    reportNumber = "BIHEP-TH-2009-004, CERN-OPEN-2009-010, LAL-09-115",
    doi = "10.1140/epjc/s10052-010-1246-1",
    journal = "Eur. Phys. J. C",
    volume = "66",
    pages = "1--9",
    year = "2010"
}

@article{PhysRevLett.132.231903,
  title = {Measurement of the Pion Form Factor with CMD-3 Detector and Its Implication to the Hadronic Contribution to Muon ($g\ensuremath{-}2$)},
  author = {Ignatov, F. V. and others},
  collaboration = {CMD-3 Collaboration},
  journal = {Phys. Rev. Lett.},
  volume = {132},
  issue = {23},
  pages = {231903},
  numpages = {8},
  year = {2024},
  month = {Jun},
  publisher = {American Physical Society},
  doi = {10.1103/PhysRevLett.132.231903},
%  url = {https://link.aps.org/doi/10.1103/PhysRevLett.132.231903}
}

@article{PhysRevD.109.112002,
  title = {Measurement of the ${e}^{+}{e}^{\ensuremath{-}}\ensuremath{\rightarrow}{\ensuremath{\pi}}^{+}{\ensuremath{\pi}}^{\ensuremath{-}}$ cross section from threshold to 1.2 GeV with the CMD-3 detector},
  author = {Ignatov, F. V. and others},
  collaboration = {CMD-3 Collaboration},
  journal = {Phys. Rev. D},
  volume = {109},
  issue = {11},
  pages = {112002},
  numpages = {35},
  year = {2024},
  month = {Jun},
  publisher = {American Physical Society},
  doi = {10.1103/PhysRevD.109.112002},
%  url = {https://link.aps.org/doi/10.1103/PhysRevD.109.112002}
}

@article{Muong-2:2025xyk,
    author = "Aguillard, D. P. and others",
    collaboration = "Muon g-2",
    title = "{Measurement of the Positive Muon Anomalous Magnetic Moment to 127~ppb}",
    eprint = "2506.03069",
    archivePrefix = "arXiv",
    primaryClass = "hep-ex",
    reportNumber = "FERMILAB-PUB-25-0364-PPD",
    doi = "10.1103/7clf-sm2v",
    journal = "Phys. Rev. Lett.",
    volume = "135",
    number = "10",
    pages = "101802",
    year = "2025"
}

@article{Aliberti:2025beg,
    author = "Aliberti, R. and others",
    title = "{The anomalous magnetic moment of the muon in the Standard Model: an update}",
    eprint = "2505.21476",
    archivePrefix = "arXiv",
    primaryClass = "hep-ph",
    reportNumber = "CERN-TH-2025-101, FERMILAB-PUB-25-0344-T, INT-PUB-25-015, IPARCOS-UCM-25-029, KEK Preprint 2025-22, LTH 1403, MITP-25-037, UWThPh 2025-15, UWThPh
  2025-15, ZU-TH 37/25, IPARCOS-UCM-25-029",
    doi = "10.1016/j.physrep.2025.08.002",
    journal = "Phys. Rept.",
    volume = "1143",
    pages = "1--158",
    year = "2025"
}

@article{Aoyama:2020ynm,
    author = "Aoyama, T. and others",
    title = "{The anomalous magnetic moment of the muon in the Standard Model}",
    eprint = "2006.04822",
    archivePrefix = "arXiv",
    primaryClass = "hep-ph",
    reportNumber = "FERMILAB-PUB-20-207-T, INT-PUB-20-021, KEK Preprint 2020-5,
  MITP/20-028, KEK Preprint 2020-5, MITP/20-028, CERN-TH-2020-075, IFT-UAM/CSIC-20-74, LMU-ASC 18/20, LTH 1234,
  LU TP 20-20, LTH 1234, LU TP 20-20, MAN/HEP/2020/003, PSI-PR-20-06, UWThPh 2020-14, ZU-TH 18/20",
    doi = "10.1016/j.physrep.2020.07.006",
    journal = "Phys. Rept.",
    volume = "887",
    pages = "1--166",
    year = "2020"
}

@article{dirac1928quantum,
  title={The quantum theory of the electron. Part II},
  author={Dirac, Paul Adrien Maurice},
  journal={Proceedings of the royal society of London. Series A, Containing papers of a mathematical and physical character},
  volume={118},
  number={779},
  pages={351--361},
  year={1928},
  publisher={The Royal Society London}
}
\end{document}